\documentclass[asa,12pt,psfig,twoside]{article}
\usepackage{amsmath}
\usepackage{amsthm}
\usepackage{mathptmx}
\usepackage{graphicx}
\usepackage{float}
\usepackage[titletoc,title]{appendix}
\usepackage[labelfont=bf]{caption} 
\usepackage[a4paper,margin=1in]{geometry}
\usepackage[hang,flushmargin]{footmisc}
\usepackage[%
colorlinks=true,%
urlcolor=blue%
]{hyperref}
\usepackage{apacite} 
\usepackage{sectsty}
\usepackage{breqn}

\sectionfont{\fontsize{12}{15}\selectfont}
\subsectionfont{\fontsize{12}{15}\selectfont}
\subsubsectionfont{\fontsize{12}{15}\selectfont}

\usepackage{titlesec}
\titlelabel{\thetitle.\quad}
\titlespacing\section{-5pt}{12pt plus 4pt minus 2pt}{0pt plus 2pt minus 2pt}
\titlespacing\subsection{-5pt}{12pt plus 4pt minus 2pt}{0pt plus 2pt minus 2pt}
\titlespacing\subsubsection{-5pt}{12pt plus 4pt minus 2pt}{0pt plus 2pt minus 2pt}

\def\mybibliography#1{{\noindent \Large \bf References}\list
 {}{\setlength{\leftmargin}{0.4in}\setlength{\labelsep}{0pt}
\itemindent=-\leftmargin}
 \def\newblock{\hskip .02em plus .20em minus -.07em}
 \sloppy\clubpenalty4000\widowpenalty4000
 \sfcode`\.=1000\relax}
\newbox\TempBox \newbox\TempBoxA
\def\uw#1{%
  \ifmmode\setbox\TempBox=\hbox{$#1$}\else\setbox\TempBox=\hbox{#1}\fi%
  \setbox\TempBoxA=\hbox to \wd\TempBox{\hss\char'176\hss}%
  \rlap{\copy\TempBox}\smash{\lower9pt\hbox{\copy\TempBoxA}}%
}
\newbox\TempBox \newbox\TempBoxA
\def\uwd#1{%
  \ifmmode\setbox\TempBox=\hbox{$#1$}\else\setbox\TempBox=\hbox{#1}\fi%
  \setbox\TempBoxA=\hbox to \wd\TempBox{\hss\char'176\hss}%
  \rlap{\copy\TempBox}\smash{\lower10pt\hbox{\copy\TempBoxA}}%
}
\def\mathunderaccent#1{\let\theaccent#1\mathpalette\putaccentunder}
\def\putaccentunder#1#2{\oalign{$#1#2$\crcr\hidewidth
\vbox to.2ex{\hbox{$#1\theaccent{}$}\vss}\hidewidth}}

\usepackage{setspace}
\usepackage{indentfirst} 

\usepackage{fancyhdr}
\fancypagestyle{firstpage}{%
	\fancyhf{}%
	\setlength{\headsep}{1pt}
	\lhead{\textit{Statistics and Application} {\bf{\{ISSN 2452-7395(online)\}}}\\
	Volume .., Nos. …, … (New Series), pp ...‒…}
	
}

\usepackage{lipsum}

\newcommand\blfootnote[1]{%
	\begingroup
	\renewcommand\thefootnote{}\footnote{#1}%
	\addtocounter{footnote}{-1}%
	\endgroup
}

\newcommand{\titlesize}{\fontsize{16pt}{20pt}\selectfont}
\newcommand{\namesize}{\fontsize{12pt}{20pt}\selectfont}
\renewenvironment{abstract}
{\begin{quote}
		\noindent \rule{\linewidth}{.5pt}\par{\bfseries \abstractname}}
	{\medskip\noindent \rule{\linewidth}{.5pt}
	\end{quote}

} 

\begin{document}

\newtheorem{theorem}{Theorem}[section]

\newtheorem{proposition}{Proposition}[section]
\newtheorem{corollary}{Corollary}[section]
\newtheorem{lemma}{Lemma}[section]

\vspace*{.05in}

\begin{center}

{\titlesize \textbf{Bayesian Predictive Inference For Finite Population Quantities
\\ \vspace{6pt}
		Under Informative Sampling }}

\vspace*{12pt}
{\namesize \bf Junheng Ma\blfootnote{Corresponding Author: Balgobin Nandram\\E-mail:\href{mailto:balnan@wpi.edu}{balnan@wpi.edu} }, Joe Sedransk, Balgobin Nandram and Lu Chen}\\
\textit{Case Western Reserve University, Cleveland, OH, USA\\ Worcester Polytechnic Institute, Worcester, MA, USA}\\
\vspace*{12pt}
Received ….; Revised ….; Accepted …..
\end{center}

\begin{abstract}
		
	\setlength{\parindent}{2em}
  We investigate Bayesian predictive inference for finite population quantities when there are unequal
  probabilities of selection. Only limited information about the sample design is available;
  i.e., only the first-order selection probabilities corresponding to the sample units are known. Our
  methodology, unlike that of Chambers, Dorfman and Wang (1998), can be used to make inference
  for finite population quantities and provides measures of precision and intervals. Moreover, our
  methodology, using Markov chain Monte Carlo methods, avoids the necessity of using asymptotic closed form
  approximations, necessary for the other approaches that have been proposed. A set of simulated
  examples shows that the informative model provides improved precision over a standard ignorable
  model, and corrects for the selection bias.
	
	\medskip
	
	\noindent {\textit{Keywords}}: Gibbs sampler, Poisson sampling, Restricted inference, Selection bias, Selection probabilities,
	SIR algorithm, Transformation.
	
\end{abstract}

\setcounter{page}{1}
\thispagestyle{firstpage}

\newpage

\section{Introduction}
\label{S:1}

%\singlespacing
Over the last 20 years there has been considerable research about model based inference for
finite population quantities when there is a selection bias. Most of this research is summarized
in Pfeffermann and Sverchkov (2009). Our work is patterned after that of Chambers, Dorfman
and Wang (1998), hereafter CDW, who assumed that the only information about the survey design
available to the survey analyst is the set of first-order inclusion probabilities for the sampled units.
CDW noted that ``it is almost impossible to proceed without fixing ideas on an example". The example, which
they used, is a generalization of one presented by Krieger and Pfeffermann (1992) and we modify
it further to make it more plausible for applications. Note that CDW made inference only for superpopulation parameters rather than finite population quantities as we do.

The purpose of our paper is to demonstrate, by example, the value of using Bayesian methods
in complicated sample survey situations such as this one, i.e., where there is a selection bias and
limited sample information. While completely general solutions to such problems are not available
because of differences in the assumptions, our specification should be close to those seen in many
surveys. For example, in establishment surveys the selection probability is often proportional to a
measure of size which is linearly related to the variable of interest, Y. Also, the distribution of Y is
positively skewed, thereby motivating a logarithmic transformation of Y.

For the theoretical development the sample units are assumed to be chosen by Poisson sampling,
as CDW did. For the numerical examples, though, the more conventional systematic pps sampling
method is used; see Section 3. Let $\tilde{I} = (I_1,..., I_N)$ denote $N$ independent Bernoulli random variables
where $I_i = 1$ if unit $i$ is selected in the sample and $I_i = 0$, otherwise. Specifically, CDW assumed
\begin{equation}
	Pr(\tilde{I}|\pi_1,...,\pi_N)=\left( \prod_{i \in s}^{}\pi_i\right) \left( \prod_{i \notin s}^{}(1-\pi_i)\right) ,
\end{equation}
where
\begin{equation}
	\pi_i=Pr(I_i=1)=\frac{n\nu_i}{\sum_{j=1}^{N}\nu_j},
\end{equation}
Also given $\beta_0, \beta_1, \sigma_e^2$ and the $Y_i$, 
\begin{equation}
\nu_i=\beta_0+\beta_1Y_i+\epsilon_i, \qquad i=1,...,N
\end{equation}
with, independently, $\epsilon_i|\sigma_e^2 \sim N(0, \sigma_e^2)$. In (2), $(\nu_1,...,\nu_n)$, corresponding to the sampled units, are known prior to sampling. 

While (3) is a realistic specification for many establishment surveys
with $\nu_i$ a measure of size of unit $i$, in other surveys a heavy tailed distribution for the $\epsilon_i$ may be more appropriate. While CDW assumed that the $Y_i$ follow a normal distribution, we take
\begin{equation}
\log(Y_i) \sim N(\mu, \sigma^2).
\end{equation}

Following CDW we assume that the data available to the analyst are the vector of sampled values,
$y_s$, the vector of $\pi_i$ corresponding to the sampled units and $\tilde{I} = (I_1,..., I_N)$. From (2) it is clear that
$\nu_1,..., \nu_N$ are not identifiable. While many restrictions are possible, taking $\sum_{j=1}^{N}\nu_j=t$, say, is convenient and this population sum is commonly available. 

Our objective is a fully Bayesian analysis yielding exact inferences about any finite population
quantity of interest, e.g., finite population quantiles. The analysis here is complicated because, from
(2), $0 \leq \nu_i \leq t/n, \ i=n+1,..., N$ (i.e., for the non-sampled $\nu_i$) and we must make inference for
the non-sampled $\nu_i$ subject to the restriction that $\sum_{i \notin s}^{}\nu_i =t-\sum_{i \in s}^{}\nu_i$.

For convenience, we let the units indexed by $ \left\lbrace1,2, ...,n\right\rbrace   $ denote the sampled units and the
units indexed by $ \left\lbrace n+1,n+2, ...,N\right\rbrace   $ the non-sampled units. Thus, we have $I_i = 1$ for $i = 1,2,...,n$ and
$I_i = 0$ for $i = n+1, n+2,...,N$. The selection probabilities, $\left\lbrace \pi_i: i \in s \right\rbrace =\left\lbrace \pi_1, \pi_2,..., \pi_n\right\rbrace$ , and the values, $\left\lbrace y_i: i \in s \right\rbrace =\left\lbrace y_1, y_2,..., y_n\right\rbrace$, for the sampled units are assumed known and denoted
by $\pi_s$ and $\tilde{y}_s$. We use $\pi_{ns}$ and $\tilde{y}_{ns}$ to denote the selection probabilities and response values for the
non-sampled units. The vector$\left\lbrace \nu_1, \nu_2,..., \nu_N\right\rbrace$ is denoted by $\tilde{\nu}$. Also $\tilde{\psi}$ and $\tilde{\eta}$ denote the parameters $(\mu, \sigma^2)$ in (4), and $(\beta_0, \beta_1, \sigma_e^2)$ in (3), respectively.

We transform $\tilde{\nu}=\left\lbrace \nu_1, \nu_2,..., \nu_N\right\rbrace$ to $\tilde{Z}=\left\lbrace Z_1, Z_2,..., Z_N\right\rbrace$ so that the $Z_i$’s are centered at 0, $$Z_1=\nu_1-\bar{\nu}, \quad  Z_2=\nu_2-\bar{\nu}, \quad ..., \quad Z_{N-1}=\nu_{N-1}-\bar{\nu}, \quad Z_N=\bar{\nu} $$
where $\bar{\nu}=\frac{1}{N}\sum_{j=1}^{N}\nu_j=\frac{1}{N}t$.

Since $\left\lbrace \pi_i:i=1,2, ..., n\right\rbrace$ and $t$ are known, it is clear from (2) that  $\tilde{\nu}_s =(\nu_1, \nu_2,..., \nu_n )$ is
known. The size measures for the non-sampled units, $\left\lbrace \nu_{n+1}, \nu_{n+2},..., \nu_N\right\rbrace$, denoted by $\tilde{\nu}_{ns}$, are
not known at the estimation stage. From the transformation from $\tilde{\nu}$ to $\tilde{Z}$, $(z_1, z_2, ...,z_n, z_N)$ are known and denoted by $\tilde{z}_s$
while $(z_{n+1}, z_{n+2},..., z_{N-1})$ are unknown and denoted by $\tilde{z}_{ns}$.

Most of the research in this area is frequentist, well summarized in Pfeffermann and Sverchkov
(2009), often using approximations, and limited to inferences about finite population means and
totals. There are four relevant papers using Bayesian methods. To incorporate selection bias, Malec,
Davis and Cao (1999) used a hierarchical Bayes method to estimate a finite population mean when
there are binary data. Difficulty in including the selection probabilities directly in the model forces
them to make an ad hoc adjustment to the likelihood function and use a Bayes, empirical Bayes (i.e.,
not a full Bayesian) approach. Nandram and Choi (2010) and Nandram, Bhatta, Bhadra and Shen (2013) extended the Malec,
Davis and Cao (1999) model. Pfeffermann et al. (2006) assumed a more complex design than we do,
i.e., two level modeling with informative probability sampling at each of the two levels. For the most
part they used a conditional likelihood, but  presented the methodology (Section 3.2) and an example
where they used the full likelihood, but made inference only for the “super-population” parameters.
Pfeffermann et al. (2006) assumed that the covariates are known for all units in the population,
thus greatly simplifying their analysis. The two differences, i.e., assuming limited information and
making inferences for the non-sampled covariates, provide a challenging computational problem.
However, once solved, as in our paper, the methodology can be applied.

The work in this paper is based on methodology developed in Ma (2010). Recently, Zangeneh
and Little (2015) provide a different approach to a related problem of inference for the finite population
total of Y when sampling is with probability proportional to size X. They use a Bayesian
bootstrap to make inference for the values of X associated with the nonsampled units, taking account
of the assumed known population total of X. Given X, they model Y using penalized splines.
The use of the bootstrap avoids parametric assumptions, but assumes, perhaps unreasonably, that
only the sampled values of X occur in the population. Their inferential approach, reversing the usual
factorization of the distributions of $\tilde{I}$ and corresponding $\tilde{X}$ , leads to a dependence, of unknown
importance, between the parameters associated with the selection effect and the sampling distribution
of $\tilde{X}$. For a similar method based on poststratification, see Si, Pillai and Gelman (2015). 

In Section 2 we describe the Bayesian methodology while in Section 3 we use simulated examples
to compare informative sampling with ignorable sampling and with standard design based
methodology based on the Horvitz-Thompson estimator. There is further discussion in Section 4.

\section{Methodology for Informative Sampling}
\label{S:2}
\vspace*{-5pt}
We describe the Bayesian model and inference in Section 2.1, and the computational methods in Section 2.2.
\subsection{Model for Informative Sampling and Inference}
We have observed $\tilde{I}$ and $\tilde{y}_s$, the vector of sampled $Y_i$. In addition, $\tilde{z}_s$ is known. The posterior
distribution for $\tilde{y}_{ns}, \tilde{z}_{ns},  \tilde{\psi}$ and $\tilde{\eta}$ can be written as
$$\pi(\tilde{y}_{ns},\tilde{z}_{ns}, \tilde{\psi}, \tilde{\eta}|\tilde{y}_{s},\tilde{z}_{s}, \tilde{I}) \ \propto \ \pi(\tilde{y},\tilde{z}, \tilde{\psi}, \tilde{\eta}, \tilde{I}), $$
where $\pi(\tilde{y},\tilde{z}, \tilde{\psi}, \tilde{\eta}, \tilde{I})=P(\tilde{I}|\tilde{y},\tilde{z}, \tilde{\psi}, \tilde{\eta})P(\tilde{z}|\tilde{y}, \tilde{\psi}, \tilde{\eta})P(\tilde{y}| \tilde{\psi}, \tilde{\eta})P( \tilde{\psi}, \tilde{\eta})$, and $\tilde{y} \ $ and $\ \tilde{z}$ are the vectors of $Y_i$, and $\ Z_i$ for the N units in the finite population, and $\tilde{z}_{ns} \in R$,
\begin{align*}
	R = &\left\{
	(z_{n+1},z_{n+2},...,z_{N-1})|-\frac{t}{N} \leq z_i \leq \frac{t}{n}-\frac{t}{N} \ \mathrm{for} \ i=n+1,n+2,..., N-1; \right. \\\
	&\phantom{=\;\;}\left.\ \frac{t}{N}-\frac{t}{n}-\sum_{j=1}^{n}z_j \leq \sum_{j=n+1}^{N-1}z_j \leq \frac{t}{N}-\sum_{j=1}^{n}z_j \right\} 
\end{align*}

From (4), we have $$P(\tilde{I}|\tilde{y},\tilde{z}, \tilde{\psi}, \tilde{\eta})=\Big(\frac{n}{Nz_N}\Big)^n \prod_{i=1}^{n}(z_i+z_N)\prod_{i=n+1}^{N-1}\Big[1-\frac{n}{Nz_N}(z_i+z_N)\Big] \times \Big[1-\Big(\frac{n}{Nz_N}\Big)\Big(z_N-\sum_{i=1}^{N-1}z_i\Big)\Big].$$

In Appendix A we have shown that the density of $\tilde{z}$ conditional on  $\tilde{y},  \tilde{\psi}, $ and $\tilde{\eta}$ is 
\begin{align*}
P(\tilde{z}|\tilde{y}, \tilde{\psi}, \tilde{\eta}) = & \frac{1}{C(\tilde{y}_N, \tilde{\psi},\tilde{\eta} )}N(\sqrt{2\pi}\sigma_e)^{-N} \exp \left\{ -\frac{1}{2\sigma_e^2} \Bigg [ \frac{t^2}{N}-2\frac{t}{N}\sum_{i=1}^{N}\theta_i
 \right. \ \\
&\phantom{=\;\;}\left.\ +\Big(\sum_{i=1}^{N-1}z_i+\theta_N \Big)^2 + \sum_{i=1}^{N-1} (z_i-\theta_i)^2 \Bigg ] \right\}, 
\end{align*}
where $\theta_j=\beta_0+\beta_1y_j, j=1,...,N$ and $C(\tilde{y}_N,\tilde{\psi}, \tilde{\eta})$ is the normalization constant. The density of $y$ conditional on $\tilde{\psi}$ and $\tilde{\eta}$ is straightforward,
$$P(\tilde{y}| \tilde{\psi}, \tilde{\eta})= (\sqrt{2\pi}\sigma)^{-N}\exp \left\lbrace -\frac{1}{2\sigma^2}\sum_{i=1}^{N}(\log{(y_i)}-\mu)^2\right\rbrace $$
and $P(\tilde{\psi}, \tilde{\eta})$ is the prior distribution. For $(\tilde{\psi}, \tilde{\eta}),$ we use the non-informative prior distribution, i.e., $P(\tilde{\psi}, \tilde{\eta}) \propto \sigma^{-2} \sigma_e^{-2}$.

\subsection{Computational Methods}
For convenience, the posterior distribution can be further rewritten as
\begin{equation}
	\pi(\tilde{y}_{ns},\tilde{z}_{ns}, \tilde{\psi}, \tilde{\eta}|\tilde{y}_{s},\tilde{z}_{s}, \tilde{I}) \ \propto \frac{1}{C(\tilde{y}_N, \tilde{\psi}, \tilde{\eta})}\pi_a(\tilde{y}_{ns},\tilde{z}_{ns}, \tilde{\psi}, \tilde{\eta}|\tilde{y}_{s},\tilde{z}_{s}, \tilde{I})
\end{equation}
with $\tilde{z}_{ns} \in R $ and
\begin{align*}
\pi_a&(\tilde{y}_{ns},\tilde{z}_{ns}, \tilde{\psi}, \tilde{\eta}|\tilde{y}_{s},\tilde{z}_{s}, \tilde{I})=L(\sigma_e\sigma)^{-N-2} \exp \left\lbrace -\frac{1}{2\sigma^2}\sum_{i=1}^{N}(\log{(y_i)}-\mu)^2\right\rbrace \\
&\exp\left\lbrace  -\frac{1}{2\sigma_e^2} \Bigg [ \frac{t^2}{N}-2\frac{t}{N}\sum_{i=1}^{N}\theta_i+\Big(\sum_{i=1}^{N-1}z_i+\theta_N \Big)^2 + \sum_{i=1}^{N-1} (z_i-\theta_i)^2 \Bigg ] \right\rbrace, 
\end{align*}
where
$$L=\prod_{i=n+1}^{N-1}\Bigg[1-\frac{n}{Nz_N}(z_i+z_N)\Bigg]  \Bigg[1-\Big(\frac{n}{Nz_N}\Big)\Big(z_N-\sum_{i=1}^{N-1}z_i\Big)\Bigg],$$

We use the sampling importance resampling (SIR) algorithm (Smith and Gelfand 1992) and the
Gibbs sampler (Gelfand and Smith 1990) to perform the computation. The Gibbs sampler is used to
draw samples from $\pi_a(\tilde{y}_{ns},\tilde{z}_{ns}, \tilde{\psi}, \tilde{\eta}|\tilde{y}_{s},\tilde{z}_{s}, \tilde{I})$ and the SIR algorithm is used to subsample the Gibbs
sample to get a sample from the posterior distribution, $\pi(\tilde{y}_{ns},\tilde{z}_{ns}, \tilde{\psi}, \tilde{\eta}|\tilde{y}_{s},\tilde{z}_{s}, \tilde{I})$ . The Gibbs sampler is
described in Appendix B.

Letting $\Omega^{(k)}=(\tilde{y}_{ns}^{(k)},\tilde{z}_{ns}^{(k)}, \tilde{\psi}^{(k)}, \tilde{\eta}^{(k)}), k=1,2,...,M$ where M is the number of iterates obtained
from the Gibbs sampler for draws made from $\pi_a(\tilde{y}_{ns},\tilde{z}_{ns}, \tilde{\psi}, \tilde{\eta}|\tilde{y}_{s},\tilde{z}_{s}, \tilde{I})$, the weights in the SIR algorithm
are $w_k=\tilde{w}_k/\sum_{k'=1}^{M}\tilde{w}_{k'}, k=1,2,...,M$, where
 $$ \tilde{w}_k=\frac{\pi(\tilde{y}_{ns}^{(k)},\tilde{z}_{ns}^{(k)}, \tilde{\psi}^{(k)}, \tilde{\eta}^{(k)}|\tilde{y}_{s},\tilde{z}_{s}, \tilde{I})}{\pi_a(\tilde{y}_{ns}^{(k)},\tilde{z}_{ns}^{(k)}, \tilde{\psi}^{(k)}, \tilde{\eta}^{(k)}|\tilde{y}_{s},\tilde{z}_{s}, \tilde{I})}.$$
By (5), $\tilde{w} \propto 1/C(\tilde{y}_N^{(k)},\tilde{\psi}^{(k)}, \tilde{\eta}^{(k)}), k=1,2,...,M.$ Thus, the SIR algorithm is performed by drawing
$M_0$ iterates from $(\Omega^{(k)}, w_k), k=1,2,...,M$ without replacement.

When the SIR algorithm is executed, the normalization constant $C(\tilde{y}_N, \tilde{\psi}, \tilde{\eta})$ needs to be evaluated.
One can compute $C(\tilde{y}_N, \tilde{\psi}, \tilde{\eta})$ by drawing samples from the multivariate normal density, and
counting how many samples fall in a region $R_0$, defined in Appendix A. This procedure performs poorly because when a single
value of $(z_1,z_2,...,z_{N-1})$ is drawn from the multivariate normal density, at least one restriction
for $R_0$ is not satisfied. Hence, the estimate of $C(\tilde{y}_N, \tilde{\psi}, \tilde{\eta})$ can be zero which is not desirable. We get
around this difficulty by splitting $R_0$ into two regions and converting this problem into a classical
high-dimensional integration problem and a multivariate normal probability problem. The details of
evaluating $C(\tilde{y}_N, \tilde{\psi}, \tilde{\eta})$ are given in Appendix C.

\section{Simulation Study}
In this section we compare the methodology presented in Section 2, NIG (non-ignorable), with
inferences assuming an ignorable model (IG) and standard design based methodology using the
Horvitz-Thompson (HT) estimator. For IG the model is given by (2) and (3), i.e., without any consideration of the selection probabilities. For the standard design based inference we use the Horvitz-Thompson point estimator, $\hat{Y} =
\frac{t}{n}\sum_{i=1}^{n}(y_i/\nu_i)$. (Recall that $\sum_{i=1}^{N} \nu_i = t$ is known.) The standard $100(1-\alpha)$ percent interval is $\hat{Y} \pm  z_{\alpha/2}\frac{1}{n(n-1)}\sum_{i=1}^{n}(\frac{y_i}{p_i}-\hat{Y})^2$ where $p_i=\nu_i/t$. 

Here we emphasize inference for the finite population total, i.e.,
point estimation and nominal 95 percent intervals. Inference for finite population quantiles and other
quantities can be made in a straightforward manner: use (5) and the methodology in the Appendices to make inference for $\tilde{y}_{ns}$, then for $\tilde{y}$.

We start by choosing values for the super-population parameters, i.e., $\mu, \sigma^2, \beta_0, \beta_1,$ and $\sigma_e^2$. Then we draw a finite population of size $N$ from the joint distribution of $y$ and $\nu$. From this finite population
we draw a sample of size $n$ using systematic pps sampling. We repeat these steps $K$ times. The examples presented in this section use $N = 100, \ n = 10 $ and $K = 200$. We have also used $n = 20$
and larger values of $K$ but have seen no qualitative differences in our results.

For NIG the methodology is described in Section 2.2. In this study both HPD and equal tailed
intervals were used. One thousand Gibbs samples were selected from the approximate posterior distribution $\pi_a$
(after a burn-in of 5000 runs), and 200 without replacement samples were chosen to implement the SIR algorithm.
We have also used larger numbers of Gibbs samples and SIR subsamples, but have seen no qualitative
differences in our results.

\begin{figure}[htb]
	\begin{center}
		\caption{\textbf{Sample Means vs. Posterior Means for Non-sampled Units ($\mu=0.5, \sigma=0.16, \beta_0=0, \beta_1=1, \sigma_e=1$)}}
		\includegraphics[height=3in,width=5.2in]{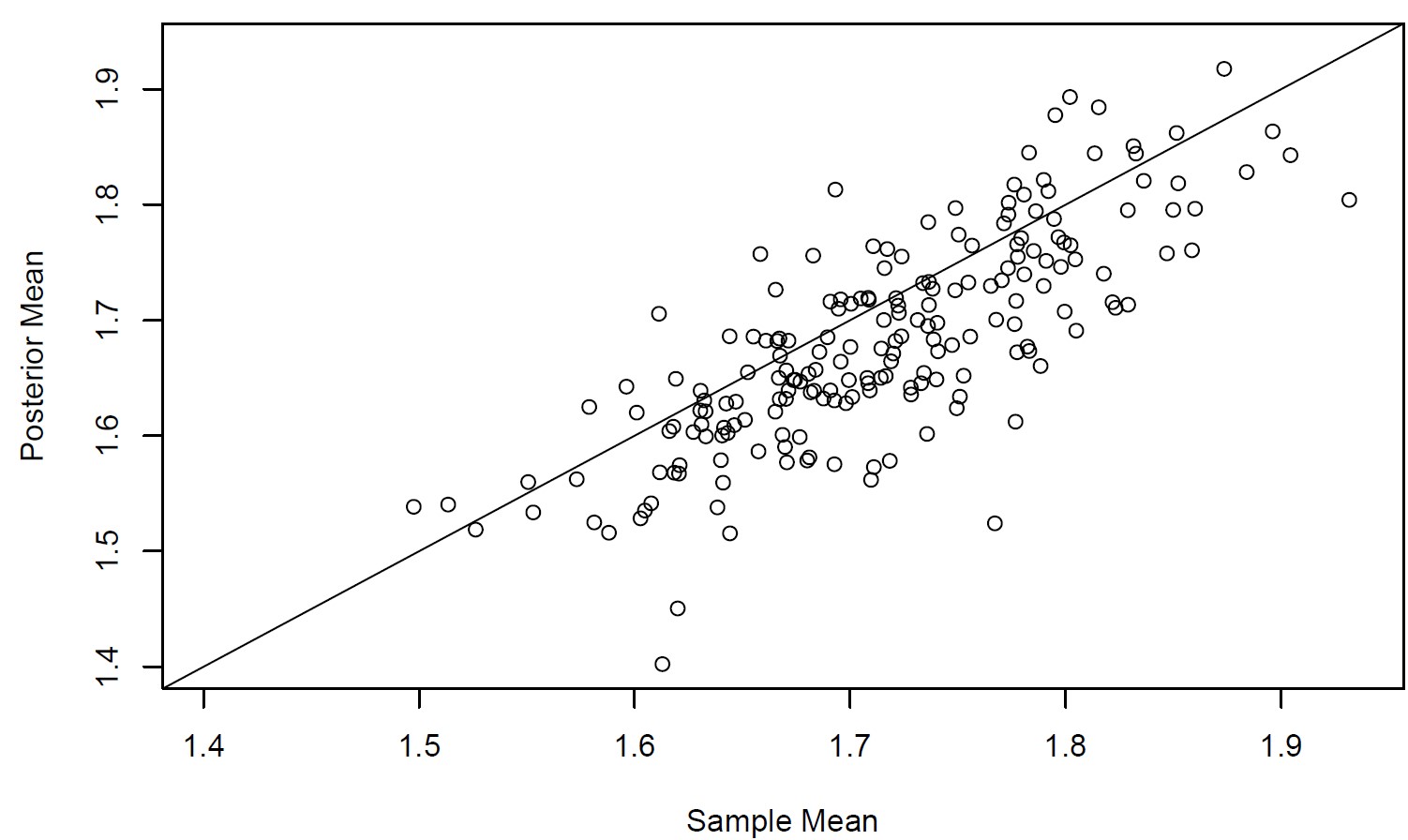}
		\label{FG:MM}
	\end{center}
\end{figure}

Our first comparisons use the relative bias, interval coverage and width, averaged over the $K$
finite populations, i.e., these are unconditional evaluations. First we plot in Figure \ref{FG:MM} the values of the
sample mean ($X$ axis) vs. the posterior mean for the non-sampled units ($Y$ axis) for NIG for 200 finite
populations selected from the super-population with $ \mu = 0.5, \sigma^2 = 0.16^2, \beta_0 = 0, \beta_1 = 1, \sigma_e^2=1.$ Clearly, as expected, the sample means tend to be large, the posterior means (for the non-sampled units) small, as they should be. 

Tables \ref{TAB:bias1} and \ref{TAB:bias2} compare NIG with IG for several choices of the super-population parameters,
chosen to yield a range of correlations between $y$ and $\nu$ but restricted to cases where $\nu$ is proportional
to $y$.
\captionsetup[table]{skip=0pt} 
\begin{table}[htb]
	\caption{\textbf{Relative Bias Comparison between NIG and IG}}
	\label{TAB:bias1}
	\centering
	\begin{tabular}{|ccc|c|c|c|c|}
		\hline
		\multicolumn{3}{|c|}{} & \multicolumn{4}{c|}{Relative Bias}                 \\ \hline
		\multicolumn{3}{|c|}{$(\beta_0, \beta_1, \sigma_e)=(0,1,1)$} & \multicolumn{2}{c|}{IG} & \multicolumn{2}{c|}{NIG} \\ \hline
		$\mu$	&  $\sigma$      &  $corr(Y,v)$     & $E(Y)$          & $\bar{Y}$          & $E(Y)$           & $\bar{Y}$          \\ \hline
		0.5   & 0.16   & 0.25  & 0.0355     & 0.0283     & 0.0106      & 0.0034     \\
		0.5   & 0.38   & 0.57  & 0.2177     & 0.1893     & 0.0159      & -0.0058    \\
		0.5   & 0.70   & 0.86  & 0.9884     & 0.8246     & -0.0049     & -0.0694    \\
		1.0   & 0.10   & 0.26  & 0.0108     & 0.0108     & 0.0008      & 0.0010     \\
		1.0   & 0.25   & 0.58  & 0.0754     & 0.0746     & 0.0050      & 0.0045     \\
		1.0   & 0.50   & 0.85  & 0.3798     & 0.3544     & -0.0123     & -0.0257    \\
		1.5   & 0.06   & 0.26  & 0.0069     & 0.0067     & 0.0027      & 0.0026     \\
		1.5   & 0.15   & 0.56  & 0.0320     & 0.0305     & 0.0060      & 0.0046     \\
		1.5   & 0.35   & 0.86  & 0.1812     & 0.1765     & 0.0034      & 0.0003     \\
		2.0   & 0.04   & 0.28  & 0.0011     & 0.0015     & -0.0004     & 0.0000     \\
		2.0   & 0.10   & 0.59  & 0.0158     & 0.0151     & 0.0058      & 0.0051     \\
		2.0   & 0.20   & 0.83  & 0.0529     & 0.0524     & 0.0040      & 0.0036     \\ \hline
	\end{tabular}
\end{table}
 Table \ref{TAB:bias1} presents the relative bias associated with $E(Y) = e^{\mu+(\sigma^2/2)}$ and the finite population
mean, $\bar{Y}$ . As expected, the relative bias for NIG is very small while that for IG is large for moderate
to large correlations. Table \ref{TAB:bias2} compares the average widths and coverages for NIG and IG. Clearly
for moderate to large correlations the average widths of the intervals for IG are much too large, e.g.,
4.0554 for IG vs. 0.9287 for NIG (see $\mu=1.0, \sigma=0.50$; inference for $\bar{Y}$). To make the two methods
comparable we adjusted the width of each IG interval to make it the same as the corresponding NIG
interval and used these intervals to evaluate the adjusted coverage of IG in the last column. For example, for $\mu=2.0, \sigma=0.20$ and $E(Y)$, the widths are quite different for IG(2.3391) and NIG(1.2768). Making the width for IG equal to that for NIG, 1.2768, the coverage for IG is 0.7300 (column 6) which should be compared with 0.9600 (column 11), the coverage for NIG.
\captionsetup[table]{skip=0pt} 
\begin{table}[htb]
	\centering
	\caption{\textbf{CI Width and Coverage Probability Comparison between NIG and IG}}
	\label{TAB:bias2}
	\resizebox{\textwidth}{!}{\begin{tabular}{|ccc|ccc|ccc|cc|cc|}
			\hline
			&                          &                           & \multicolumn{10}{c|}{95\% CI}                                                                                                                                                                                                                                                  \\ \hline
			\multicolumn{3}{|c|}{$(\beta_0, \beta_1, \sigma_e)=(0,1,1)$}                                                          & \multicolumn{6}{c|}{IG}                                                                                                                                                       & \multicolumn{4}{c|}{NIG}                                                                       \\ \hline
			$\mu$                        & $\sigma$                        & $corr(Y,v)$                      & \multicolumn{3}{c|}{E(Y)}                                                             & \multicolumn{3}{c|}{$\bar{Y}$}                                                                & \multicolumn{2}{c}{E(Y)}                                 & \multicolumn{2}{c|}{$\bar{Y}$}              \\ \hline
			\multicolumn{3}{|c|}{}                                                          & width                      & CP                         & adjusted CP                 & width                      & CP                         & adjusted CP                 & width                      & CP                          & width                      & CP     \\ \hline
			0.5                      & 0.16                     & 0.25                      & 0.3961                     & 0.9400                     & 0.9200                      & 0.3753                     & 0.9300                     & 0.9100                      & 0.3488                     & 0.9050                      & 0.3301                     & 0.9400 \\
			0.5                      & 0.38                     & 0.57                      & 1.5487                     & 0.8850                     & 0.5400                      & 1.3378                     & 0.9000                     & 0.5400                      & 0.7681                     & 0.9150                      & 0.7076                     & 0.9100 \\
			0.5                      & 0.70                     & 0.86                      & 10.6780                    & 0.9650                     & 0.0150                      & 8.7670                     & 0.9700                     & 0.0050                      & 1.0286                     & 0.9500                      & 0.7913                     & 0.8100 \\
			1.0                      & 0.10                     & 0.26                      & 0.3679                     & 0.9150                     & 0.8600                      & 0.3489                     & 0.9300                     & 0.8700                      & 0.3295                     & 0.8850                      & 0.3092                     & 0.8850 \\
			1.0                      & 0.25                     & 0.58                      & 1.0931                     & 0.9450                     & 0.7750                      & 1.0354                     & 0.9400                     & 0.7900                      & 0.7685                     & 0.9350                      & 0.7128                     & 0.9200 \\
			1.0                      & 0.50                     & 0.85                      & 4.3844                     & 0.9700                     & 0.1250                      & 4.0554                     & 0.9600                     & 0.0750                      & 1.1294                     & 0.9400                      & 0.9287                     & 0.9100 \\
			1.5                      & 0.06                     & 0.26                      & 0.3771                     & 0.9350                     & 0.9150                      & 0.3577                     & 0.9450                     & 0.9100                      & 0.3351                     & 0.9150                      & 0.3142                     & 0.9150 \\
			1.5                      & 0.15                     & 0.56                      & 0.9624                     & 0.9500                     & 0.8350                      & 0.9121                     & 0.9550                     & 0.8200                      & 0.7122                     & 0.9050                      & 0.6501                     & 0.9200 \\
			1.5                      & 0.35                     & 0.86                      & 3.4085                     & 0.9600                     & 0.3450                      & 3.2234                     & 0.9700                     & 0.2800                      & 1.2770                     & 0.9250                      & 1.0563                     & 0.9450 \\
			2.0                      & 0.04                     & 0.28                      & 0.4136                     & 0.9500                     & 0.8750                      & 0.3923                     & 0.9350                     & 0.8850                      & 0.3524                     & 0.8800                      & 0.3286                     & 0.8750 \\
			2.0                      & 0.10                     & 0.59                      & 1.0556                     & 0.9750                     & 0.9150                      & 1.0015                     & 0.9700                     & 0.8800                      & 0.7795                     & 0.9350                      & 0.7082                     & 0.9100 \\
			2.0   & 0.20   & 0.83  & 2.3391  & 1.0000 & 0.7300      & 2.2170 & 0.9800 & 0.6450      & 1.2768      & 0.9600     & 1.0809     & 0.8950   \\ \hline
	\end{tabular}}
\end{table}

That is, for either $E(Y)$ or $\bar{Y}$ compare the values of “adjusted CP” for IG with “CP” for NIG. For
small corr$(Y,\nu)$ the two coverages are similar. However, as expected, for moderate to large values of corr$(Y,\nu)$
the coverage for NIG is generally close to the nominal 0.95 while that for IG is smaller, markedly
so for the large correlations.

In Table \ref{TAB:bias3} we compare NIG with the standard design based method using the same set of superpopulation
parameters as in Tables \ref{TAB:bias1} and \ref{TAB:bias2}. The relative bias of each method is small, as expected.
\captionsetup[table]{skip=0pt} 
\begin{table}[htb]
	\centering
	\caption{\bf Relative Bias and CI Comparison between NIG and HT for $\beta_0=0$}
	\label{TAB:bias3}
	\resizebox{\textwidth}{!}{\begin{tabular}{|ccc|cc|ccc|cc|}
		\hline
		\multicolumn{3}{|c|}{} & \multicolumn{2}{c|}{Relative Bias} & \multicolumn{5}{c|}{95\% CI}                             \\ \hline
		\multicolumn{3}{|c|}{$(\beta_0, \beta_1, \sigma_e)=(0,1,1)$} & HT              & NIG             & \multicolumn{3}{c|}{HT}       & \multicolumn{2}{c|}{NIG} \\ \hline
		$\mu$      & $\sigma$     & $corr(Y,v)$  & $\bar{Y}$               & $\bar{Y}$                & width  & CP     & adjusted CP & width       & CP         \\ \hline
		0.50   & 0.16  & 0.25  & -0.0312         & 0.0037          & 1.0936 & 0.9700 & 0.5650      & 0.3295      & 0.9400     \\
		0.50   & 0.38  & 0.57  & -0.0335         & -0.0055         & 1.0027 & 0.9150 & 0.8250      & 0.7082      & 0.9100     \\
		0.50   & 0.70  & 0.86  & -0.0084         & -0.0694         & 1.1849 & 0.8950 & 0.8150      & 0.7915      & 0.7950     \\
		1.00   & 0.10  & 0.26  & -0.0210         & 0.0010          & 1.2400 & 0.9900 & 0.5700      & 0.3087      & 0.8800     \\
		1.00   & 0.25  & 0.58  & -0.0166         & 0.0045          & 1.2630 & 0.9400 & 0.8050      & 0.7099      & 0.9250     \\
		1.00   & 0.50  & 0.85  & -0.0181         & -0.0261         & 1.1703 & 0.8850 & 0.8850      & 0.9275      & 0.8950     \\
		1.50   & 0.06  & 0.26  & -0.0040         & 0.0026          & 1.2794 & 1.0000 & 0.6050      & 0.3154      & 0.9250     \\
		1.50   & 0.15  & 0.56  & -0.0035         & 0.0047          & 1.2439 & 0.9950 & 0.8400      & 0.6574      & 0.9150     \\
		1.50   & 0.35  & 0.86  & -0.0045         & 0.0002          & 1.2843 & 0.9550 & 0.9150      & 1.0626      & 0.9400     \\
		2.00   & 0.04  & 0.28  & -0.0012         & 0.0000          & 1.2845 & 1.0000 & 0.6100      & 0.3291      & 0.8800     \\
		2.00   & 0.10  & 0.59  & 0.0038          & 0.0051          & 1.2869 & 1.0000 & 0.8500      & 0.7078      & 0.9150     \\
		2.00   & 0.20  & 0.83  & -0.0004         & 0.0037          & 1.2739 & 0.9500 & 0.9250      & 1.0825      & 0.9050     \\ \hline
	\end{tabular}}
\end{table}
However, for each set of parameter values, the average width of the HT interval is much wider than
its NIG counterpart, leading to somewhat better coverage. To make the two methods comparable we
adjusted the width of each HT interval to make it the same as the corresponding NIG interval and
used these intervals to evaluate the adjusted coverage of HT in the $8^{th}$ column. Thus, the coverage
for NIG is much better when there is small to moderate correlation while there is little difference
when there is a large correlation. This, too, is not surprising since the HT based interval should
perform very well when $\nu$ is proportional to $y$.

We next compare NIG with HT assuming that the intercept, $\beta_0$, is larger than zero. Table 4 gives
the relative biases and interval widths and coverages. It is clear that the biases are small, as expected. When the correlation is moderate to large, the widths of the nominal 95 percent intervals for HT are
much wider than those for NIG. Making the adjustment described above so that the widths of the
HT and NIG intervals are the same, the coverage for NIG is better than HT, markedly so when the
correlation is large, e.g., 0.94 vs. 0.21 when the correlation is 0.99 ($\beta_0 = 50, \sigma_e = 0.1$). A referee noted that a comparison with the GREG estimator may have been more appropriate when, as here, $\beta_0>0$.
\captionsetup[table]{skip=0pt} 
\begin{table}[htb]
	\centering
	\caption{\bf  Relative Bias and CI Comparison between NIG and HT for $\beta_0>0$}
	\label{my-label}
	\resizebox{\textwidth}{!}{\begin{tabular}{|ccc|cc|ccc|cc|}
		\hline
		\multicolumn{3}{|c|}{} & \multicolumn{2}{c|}{Relative Bias} & \multicolumn{5}{c|}{95\% CI}                             \\ \hline
		\multicolumn{3}{|c|}{$(\beta_0, \beta_1, \sigma_e)=(0.5, 0.7, 1)$} & HT               & NIG             & \multicolumn{3}{c|}{HT}       & \multicolumn{2}{c|}{NIG} \\ \hline
		$\beta_0$    & $\sigma_e$   & $corr(Y,v)$ & $\bar{Y}$                & $\bar{Y}$               & width  & CP     & adjusted CP & width       & CP         \\ \hline
		10   & 5.5 & 0.27      & -0.0130          & 0.0279          & 1.9767 & 0.8850 & 0.8900      & 2.0714      & 0.9100     \\
		10   & 2.5 & 0.57      & 0.0056           & 0.0439          & 1.5703 & 0.9350 & 0.9400      & 1.7180      & 0.9300     \\
		10   & 1   & 0.86      & 0.0062           & 0.0020          & 1.5234 & 0.9850 & 0.9350      & 1.0249      & 0.9500     \\
		10   & 0.1 & 0.99      & -0.0026          & -0.0016         & 1.5505 & 1.0000 & 0.2400      & 0.1336      & 0.9300     \\
		50   & 5.5 & 0.27      & -0.0249          & 0.0310          & 1.7155 & 0.8800 & 0.8950      & 1.9732      & 0.9100     \\
		50   & 2.5 & 0.57      & 0.0111           & 0.0370          & 1.8071 & 0.9100 & 0.8750      & 1.6656      & 0.9150     \\
		50   & 1   & 0.86      & -0.0125          & -0.0107         & 1.7643 & 0.9700 & 0.8150      & 1.0444      & 0.9300     \\
		50   & 0.1 & 0.99      & -0.0097          & -0.0019         & 1.8269 & 1.0000 & 0.2100      & 0.1414      & 0.9400     \\
		400  & 5.5 & 0.27      & 0.0072           & 0.0556          & 1.9680 & 0.8950 & 0.8900      & 2.0915      & 0.9000     \\
		400  & 2.5 & 0.57      & -0.0051          & 0.0198          & 1.8365 & 0.9300 & 0.9250      & 1.7082      & 0.9350     \\
		400  & 1   & 0.86      & -0.0048          & -0.0061         & 1.8423 & 0.9700 & 0.8250      & 1.0287      & 0.9350     \\
		400  & 0.1 & 0.99      & 0.0058           & -0.0008         & 1.9884 & 1.0000 & 0.2250      & 0.1402      & 0.9550     \\ \hline
	\end{tabular}}
\end{table}

\begin{figure}[!htb]
	\begin{center}
		\caption{\textbf{Relative Bias Plot for $\bar{Y}$ for NIG (filled circle) and HT (empty circle) ($\mu=0.5, \sigma=0.16, \beta_0=0, \beta_1=1, \sigma_e=1$)}}
		\includegraphics[height=3in,width=5.2in]{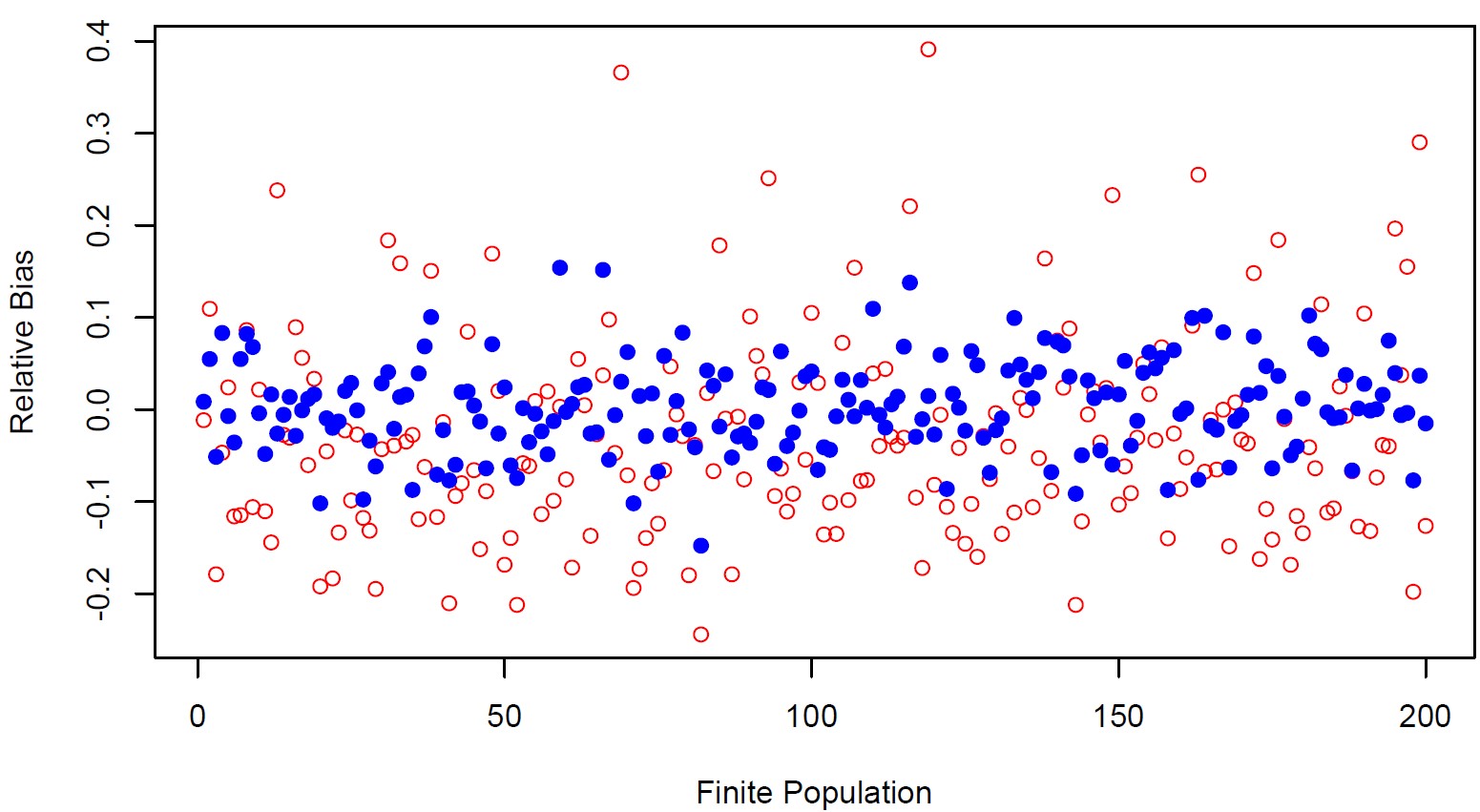}
		\label{FG:relativebias1}
	\end{center}
\end{figure}

\begin{figure}[!htb]
	\begin{center}
		\caption{\textbf{Relative Bias Plot for $\bar{Y}$ for NIG (filled circle) and HT (empty circle) ($\mu=0.5, \sigma=0.7, \beta_0=50, \beta_1=1, \sigma_e=0.1$)}}
		\includegraphics[height=3in,width=5.2in]{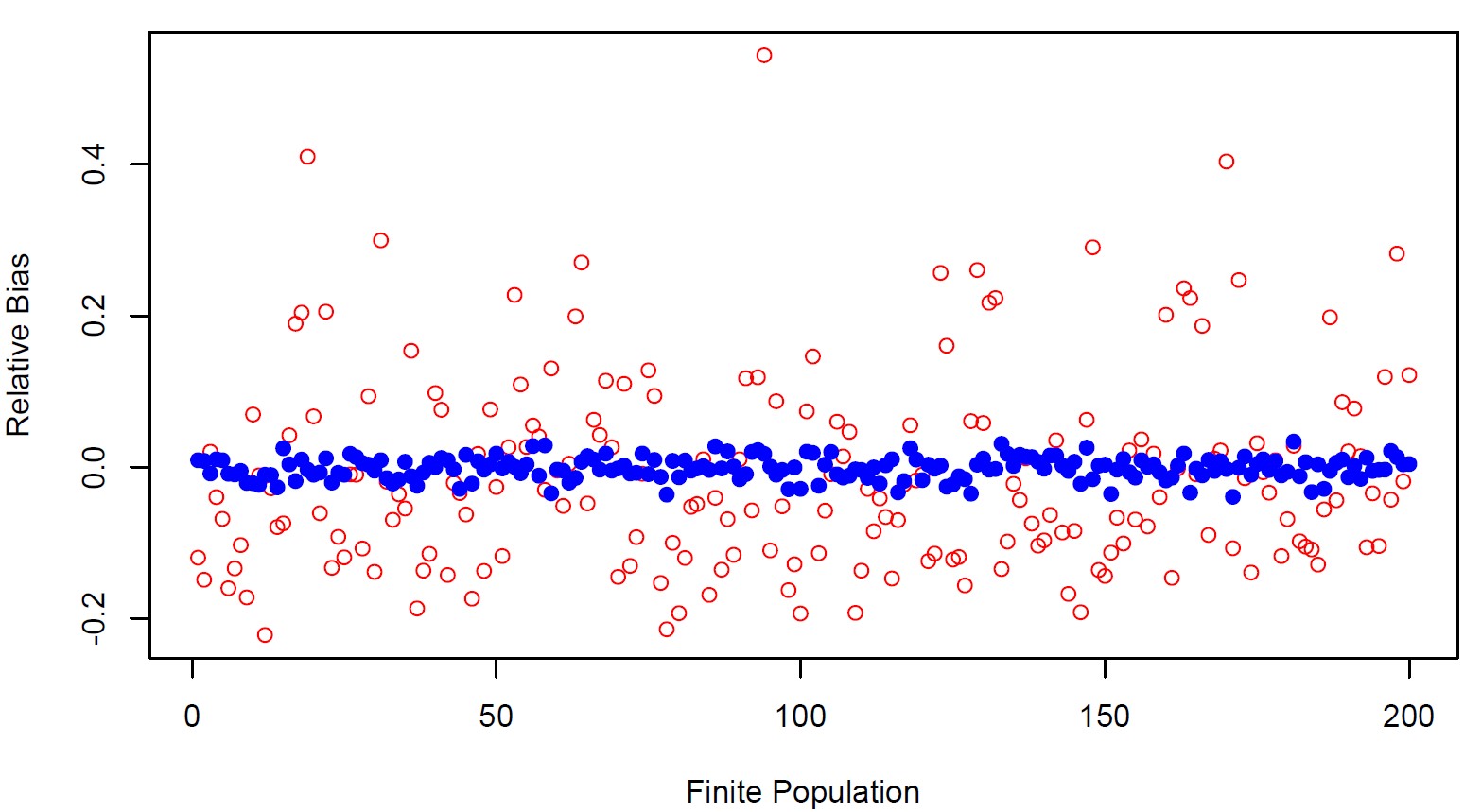}
		\label{FG:relativebias2}
	\end{center}
\end{figure}

The results presented in Tables 1-4 are averages over a large set of finite populations and do not
display variation over the populations. Figure \ref{FG:relativebias1} shows the relative bias ($Y$ axis) plotted against the
200 finite populations for NIG (filled circle) and HT (empty circle) with parameter setting: $\mu = 0.5,
 \sigma = 0.16, \beta_0 =
0, \beta_1=1, \sigma_e = 1$. This is a case with $\beta_0 = 0$, supposedly favorable to the HT estimator, but one
can see that the relative bias of HT varies more widely than the relative bias of NIG. Thus, many
individual HT samples will have a large relative bias. In Figure \ref{FG:relativebias2}, we assume an intercept greater than zero
with parameter setting: $\mu = 0.5, \sigma = 0.7, \beta_0 =
50, \beta_1=1, \sigma_e = 0.1$. Here, the relative bias of NIG
is essentially constant over the populations while that for HT varies widely. Clearly, these figures
show the advantage of using NIG, i.e., more consistent conditional performance.

Finally, note that we have considered a large number of combinations of the super-population
parameters, i.e., $\mu , \sigma, \beta_0, \beta_1, \sigma_e$. The results shown are typical of this large ensemble of evaluations.

\section{Summary}
We have shown how to carry out a complete analysis of a complicated problem using survey
data; i.e., where the analyst has only limited information about the survey design and there is a selection bias. Our model in (2) and (3) is appropriate for many establishment surveys while our specification of the model for sample selection in (4) should provide a good approximation for many survey designs.

Our examples show that relating the selection probabilities to the responses will provide more
appropriate intervals vis a vis a model that does not account for selection bias. This is especially
true when the correlation between the variable of interest, Y, and the covariate, $\nu$, is high, a situation
common in establishment surveys.

There are also (overall) gains for our method when compared with a standard approach based
on the Horvitz-Thompson (HT) estimator. Of special note is the improved conditional performance.
While the unconditional bias for the HT method may be small, it is common
to have substantial variation over the samples. Conversely, the conditional bias associated
with our method has significantly less variation (Figures 2-3).

\begin{appendices}
\section{Distribution of $\tilde{Z}$ Under Restrictions}

	Based on the transformation from $\tilde{\nu}$ to $\tilde{Z}$, the reverse transformation is 
	$$\nu_1=Z_1+Z_N,  \quad \nu_2=Z_2+Z_N, \quad , ..., \quad \nu_{N-1}=Z_{N-1}+Z_N, \quad \nu_N=Z_N-\sum_{i=1}^{N-1}Z_i.$$
	
The Jacobian of this transformation can be computed as
$$\begin{pmatrix}
	1 & 0 & 0 &... & 0 & 1 \\
	0 & 1 & 0 &... & 0 & 1 \\
	. & . & . &... & . & . \\
	0 & 0 & 0 &... & 1 & 1 \\
	-1 & -1 & -1 &... & -1 & -1 \\
\end{pmatrix}_{N \times N}$$
and $|J|=N$. For $i=1,2,..., N,$ the distribution of $\nu_i$ given $\beta_0, \beta_1, \sigma_e^2,$ and $Y_i$ is
$$\nu_i | \beta_0, \beta_1, \sigma_e^2, Y_i \sim N(\beta_0+\beta_1Y_i, \ \sigma_e^2)$$
i.e., $$f_{\nu_i}(\nu_i)=\frac{1}{\sqrt{2\pi}\sigma_e}\exp \left\lbrace -\frac{(\nu_i-\beta_0-\beta_1y_i)^2}{2\sigma_e^2}\right\rbrace. $$
Due to the fact that $\nu_i$’s are independent, the joint distribution of $\tilde{\nu}=(\nu_1,\nu_2,...,\nu_N)$ given $\beta_0, \beta_1, \sigma_e^2,$ and $\tilde{Y}$ is 
	$$f_{\tilde{\nu_i}}(\nu_1, \nu_2, ..., \nu_N)=\prod_{i=1}^{N}\Bigg[\frac{1}{\sqrt{2\pi}\sigma_e}\exp \left\lbrace -\frac{(\nu_i-\beta_0-\beta_1y_i)^2}{2\sigma_e^2}\right\rbrace\Bigg]. $$
Then the distribution of $\tilde{Z}$ given $\beta_0, \beta_1, \sigma_e^2,$ and $\tilde{Y}$ can be written as	
\begin{align*}
f_{\tilde{Z}}(z_1, z_2, ..., z_N) = & N(\sqrt{2\pi}\sigma_e)^{-N} \exp \left\{ -\frac{1}{2\sigma_e^2} \Bigg [  Nz_N^2-2z_N\sum_{i=1}^{N}\theta_i
\right. \ \\
&\phantom{=\;\;}\left.\ +\Big(\sum_{i=1}^{N-1}z_i+\theta_N \Big)^2 + \sum_{i=1}^{N-1} (z_i-\theta_i)^2 \Bigg ] \right\}, 
\end{align*}
where $\theta_j=\beta_0+\beta_1y_j$ for $j=1,2,...,N$.

There are also some restrictions for $\tilde{Z}$, which are related to the restrictions for $\tilde{\nu}$. The restrictions for $\tilde{\nu}$ can be summarized as
\begin{enumerate}
	\item $0 \leq \frac{n\nu_i}{\sum_{j=1}^{N}\nu_j} \leq 1$ for $i = 1,2,...,N$;
	\item $\sum_{j=1}^{N}\nu_j=t$
\end{enumerate}
Based on the transformation from $\tilde{\nu}$ to $\tilde{Z}$ , the restrictions for $\tilde{Z}$ are
\begin{enumerate}
	\item $z_N=\frac{t}{N}$;
	\item $-\frac{t}{N} \leq z_i \leq \frac{t}{n}-\frac{t}{N}$ for $i=1,2,...,N-1;$
	\item $\frac{t}{N}-\frac{t}{n} \leq \sum_{j=1}^{N-1} z_j \leq \frac{t}{N}.$
\end{enumerate}
Given $Z_N=\frac{t}{N}$, the conditional distribution of $(Z_1, Z_2,..., Z_{N-1})$ is 
$$f(z_1,z_2, ..., z_{N-1}|z_N=\frac{t}{N})=\frac{1}{f(z_N=\frac{t}{N})}f(z_1,z_2, ..., z_{N-1},z_N=\frac{t}{N}).$$
Then under the second and third restrictions for $\tilde{Z}$, the conditional distribution of $(z_1,z_2, ..., z_{N-1})$ is 
\begin{align*}
	f_{R_0}(z_1,z_2, ..., z_{N-1}|z_N=\frac{t}{N})&=\frac{f(z_1,z_2, ..., z_{N-1}|z_N=\frac{t}{N})}{\int_{R_0}^{}f(z_1,z_2, ..., z_{N-1}|z_N=\frac{t}{N})dz_1dz_2...dz_{N-1}}\\
	&=\frac{f(z_1,z_2, ..., z_{N-1},z_N=\frac{t}{N})}{\int_{R_0}^{}f(z_1,z_2, ..., z_{N-1},z_N=\frac{t}{N})dz_1dz_2...dz_{N-1}},
\end{align*}
where 
$$R_0 =  \left\{  (z_{1},z_{2},...,z_{N-1}) | -\frac{t}{N} \leq z_i \leq \frac{t}{n}-\frac{t}{N} \ \mathrm{for} \ i=1,2,..., N-1;\frac{t}{N}-\frac{t}{n} \leq \sum_{j=1}^{N-1}z_j \leq \frac{t}{n} \right\}.$$
Finally, the distribution of  $(Z_1, Z_2,..., Z_N)$ under the restrictions for $\tilde{Z}$ is 
\begin{align*}
f_{R_0}(z_1,z_2, ..., z_{N-1},z_N) \ = \ & \frac{1}{C(\tilde{y}_N, \tilde{\psi},\tilde{\eta} )}N(\sqrt{2\pi}\sigma_e)^{-N} \exp \left\{ -\frac{1}{2\sigma_e^2} \Bigg [ \frac{t^2}{N}-2\frac{t}{N}\sum_{i=1}^{N}\theta_i
\right. \ \\
&\phantom{=\;\;}\left.\ +\Big(\sum_{i=1}^{N-1}z_i+\theta_N \Big)^2 + \sum_{i=1}^{N-1} (z_i-\theta_i)^2 \Bigg ] \right\}, 
\end{align*}
where $Z_N=\frac{t}{N}, (Z_1, Z_2,..., Z_{N-1}) \in R_0$ and 
\begin{align*}
C(\tilde{y}_N, \tilde{\psi},\tilde{\eta} ) \ = \ & \int_{R_0}^{}N(\sqrt{2\pi}\sigma_e)^{-N} \exp \left\{ -\frac{1}{2\sigma_e^2} \Bigg [ \frac{t^2}{N}-2\frac{t}{N}\sum_{i=1}^{N}\theta_i+\Big(\sum_{i=1}^{N-1}z_i+\theta_N \Big)^2
\right. \ \\
&\phantom{=\;\;}\left.\  + \sum_{i=1}^{N-1} (z_i-\theta_i)^2 \Bigg ] \right\}dz_1dz_2...dz_{N-1}. 
\end{align*}

\section{Gibbs Sampler}

We use the Gibbs sampler to draw samples from  $\pi_a(\tilde{y}_{ns},\tilde{z}_{ns}, \tilde{\psi}, \tilde{\eta}|\tilde{y}_{s},\tilde{z}_{s}, \tilde{I})$. In order to perform the
Gibbs sampler, we need to find the conditional distributions of $\tilde{y}_{ns},\tilde{z}_{ns}, \tilde{\psi}$, and $\tilde{\eta}$ respectively given everything else.

The conditional distribution of $\tilde{y}_{ns}$, given $(\tilde{y}_{s},\tilde{z}_{N}, \tilde{\psi}, \tilde{\eta}, \tilde{I})$ is 
\begin{equation}
	P(\tilde{y}_{ns}|\tilde{y}_{s}, \tilde{z}_{N}, \tilde{\psi}, \tilde{\eta}, \tilde{I}) \propto \exp \left\lbrace T_1 \sum_{i=n+1}^{N-1} \Big(y_i+\frac{T_i}{2T_1} \Big)^2 + T_1 \Big(y_N+\frac{T_2}{2T_1} \Big)^2   \right\rbrace , 
\end{equation}
where 
$$T_1=-\frac{1}{2\sigma^2}-\frac{\beta_1^2}{2\sigma_e^2}, \quad T_2=\frac{\mu}{\sigma^2}-\frac{\beta_1(\beta_0+\sum_{i=1}^{N-1}z_i-z_N)}{\sigma_e^2},$$
$$T_i=\frac{\mu}{\sigma^2}-\frac{\beta_1(\beta_0-z_i-z_N)}{\sigma_e^2}, \ i=n+1,n+2,...,N-1.$$

From (6), we see that given $(y_{n+1},...,y_{N-1},\tilde{y}_{s}, \tilde{z}_{N}, \tilde{\psi}, \tilde{\eta}, \tilde{I})$, $y_N$ has a normal distribution with mean $-\frac{T_2}{2T_1}$ and variance $-\frac{1}{2T_1}$, i.e., $$y_N|y_{n+1},...,y_{N-1},\tilde{y}_{s}, \tilde{z}_{N}, \tilde{\psi}, \tilde{\eta}, \tilde{I} \sim N(-\frac{T_2}{2T_1},\ -\frac{1}{2T_1}).$$
Also, given $(y_{n+1},...,y_{i-1},y_{i+1},...,y_{N-1},y_N,\tilde{y}_{s}, \tilde{z}_{N}, \tilde{\psi}, \tilde{\eta}, \tilde{I})$, $y_i$ has a normal distribution with mean $-\frac{T_i}{2T_1}$ and variance $-\frac{1}{2T_1}$, i.e., $$y_i|y_{n+1},...,y_{i-1},y_{i+1},...,y_{N-1},y_N,\tilde{y}_{s}, \tilde{z}_{N}, \tilde{\psi}, \tilde{\eta}, \tilde{I} \sim N(-\frac{T_i}{2T_1}, \ -\frac{1}{2T_1})$$
for $i=n+1,n+2,...,N-1.$

The conditional distribution of $\tilde{z}_{ns}$ given $(\tilde{y}_{N}, \tilde{z}_{s}, \tilde{\psi}, \tilde{\eta}, \tilde{I})$ is 
\begin{equation}
 P(\tilde{z}_{ns}|\tilde{y}_{N}, \tilde{z}_{s}, \tilde{\psi}, \tilde{\eta}, \tilde{I}) \propto \exp \left\lbrace -\frac{1}{2\sigma_e^2}\Bigg[ \Bigg(\sum_{i=n+1}^{N-1} z_i+T_3 \Bigg)^2 + \sum_{i=n+1}^{N-1} (z_i-\theta_i )^2  \Bigg] \right\rbrace ,
\end{equation}
where $\tilde{z}_{ns} \in R$ and $T_3=\sum_{i=1}^{n}z_i+\theta_N$. Notice that the right side of (7) is the kernel of a multivariate normal distribution, $MVN(\tilde{\mu}_0, \tilde{\Sigma})$, where 
\begin{equation}
	\tilde{\mu}_0=\begin{pmatrix}
	\mu_{01} \\
	\mu_{02}  \\
	...  \\
	\mu_{0(N-n-1)} \\
	\end{pmatrix} =\begin{pmatrix}
	\frac{1}{N-n}[(N-n-1)\theta_{n+1}-\theta_{n+2}-...-\theta_{N-1}-T_3] \\
	\frac{1}{N-n}[(N-n-1)\theta_{n+2}-\theta_{n+1}-...-\theta_{N-1}-T_3]  \\
	...  \\
	\frac{1}{N-n}[(N-n-1)\theta_{N-1}-\theta_{n+1}-...-\theta_{N-2}-T_3] \\
	\end{pmatrix} _{(N-n-1) \times 1}
\end{equation}
and 
\begin{equation}
	\tilde{\Sigma}^{-1}=\frac{1}{\sigma_e^2}\begin{pmatrix}
	2 & 1 & ... & 1 \\
	1 & 2 & ... & 1 \\
	. & . & ... & . \\
	1 & 1 & ... & 2 \\
	\end{pmatrix}_{(N-n-1) \times (N-n-1)}.
\end{equation}
From (7), (8), and (9) we see that $\tilde{z}_{ns}$ given $(\tilde{y}_{N}, \tilde{z}_{s}, \tilde{\psi}, \tilde{\eta}, \tilde{I})$ has a multivariate normal distribution with mean $\tilde{\mu}$ and variance $\tilde{\Sigma}$ restricted to region $R$, i.e., $$\tilde{z}_{ns}|\tilde{y}_{N}, \tilde{z}_{s}, \tilde{\psi}, \tilde{\eta}, \tilde{I} \sim MVN(\tilde{\mu}_0,\tilde{\Sigma}),$$
where $\tilde{z}_{ns} \in R$. One may draw samples from $\tilde{z}_{ns}$ given everything else by generating random samples from the multivariate normal density, $MVN(\tilde{\mu}_0, \tilde{\Sigma})$, and only keeping the samples which fall in $R$. This procedure performs poorly because there are too many rejections. Thus, we proceed by sampling the members of $\tilde{z}_{ns}$ one at a time, i.e., for $z_i, \ i=n+1,n+2,...,N-1$ given  $(z_{n+1},...,z_{i-1},z_{i+1},...,z_{N-1},\tilde{z}_{s}, \tilde{y}_{N}, \tilde{\psi}, \tilde{\eta}, \tilde{I})$, we have 
\begin{align}
\nonumber
P(z_i| & z_{n+1},...,z_{i-1},z_{i+1},...,z_{N-1},\tilde{z}_{s}, \tilde{y}_{N}, \tilde{\psi}, \tilde{\eta}, \tilde{I}) \propto \\
\nonumber
& \exp \left\lbrace  -\frac{1}{2\sigma_e^2}
 \Big(z_i+\frac{1}{2}\Big(T_3+\sum_{j=n+1}^{N-1}z_j-z_i-\theta_i \Big)^2  \right\rbrace, 
\end{align}
where $T_{maxi} \leq z_i \leq T_{mini}$ and 
$$T_{maxi} \ = \ \max \Bigg(-z_N, \ \frac{t}{n}\Bigg(n-1-\sum_{i=1}^{n}\pi_i-\frac{n(N-n-1)}{N}\Bigg)-\sum_{j=n+1}^{N-1}z_j+z_i\Bigg),$$
$$T_{mini} \ = \ \min \Bigg(\bigg(\frac{N}{n}-1\bigg)z_N, \ \frac{t}{n}\Bigg(n-\sum_{i=1}^{n}\pi_i-\frac{n(N-n-1)}{N}\Bigg)-\sum_{j=n+1}^{N-1}z_j+z_i\Bigg).$$
Then, given $(z_{n+1},...,z_{i-1},z_{i+1},...,z_{N-1},\tilde{z}_{s}, \tilde{y}_{N}, \tilde{\psi}, \tilde{\eta}, \tilde{I}), z_i, \ i=n+1,n+2,...,N-1$, has a truncated normal distribution, i.e., 
$$z_i| z_{n+1},...,z_{i-1},z_{i+1},...,z_{N-1},\tilde{z}_{s}, \tilde{y}_{N}, \tilde{\psi}, \tilde{\eta}, \tilde{I} \sim N \Bigg( -\frac{1}{2}\Big(T_3+\sum_{j=n+1}^{N-1}z_j-z_i-\theta_i \Big) , \ \frac{1}{2}\sigma_e^2 \Bigg),$$
where $T_{maxi} \leq z_i \leq T_{mini}$.

The conditional distribution of $\tilde{\psi}$, given $(\tilde{y}_N,\tilde{z}_N, \tilde{\eta}, \tilde{I})$, is 
$$P(\tilde{\psi}|\tilde{y}_N,\tilde{z}_N, \tilde{\eta}, \tilde{I}) \propto \sigma^{-N-2} \exp \left\lbrace -\frac{1}{2\sigma^2}\sum_{i=1}^{N}(y_i-\mu)^2 \right\rbrace. $$
Then the conditional distribution of $\mu$ given $(\sigma^2,\tilde{y}_N,\tilde{z}_N, \tilde{\eta}, \tilde{I})$ is 
$$P(\mu|\sigma^2, \tilde{y}_N,\tilde{z}_N, \tilde{\eta}, \tilde{I}) \propto  \exp \left\lbrace -\frac{N}{2\sigma^2}\Big(\mu-\frac{1}{N}\sum_{i=1}^{N}y_i \Big)^2 \right\rbrace.$$
Thus, given $(\sigma^2,\tilde{y}_N,\tilde{z}_N, \tilde{\eta}, \tilde{I})$, $\mu$ has a normal distribution with mean $\frac{1}{N}\sum_{i=1}^{N}y_i$ and variance $\frac{1}{N}\sigma^2$, i.e., $$\mu|\sigma^2, \tilde{y}_N,\tilde{z}_N, \tilde{\eta}, \tilde{I} \sim N \Bigg(\frac{1}{N}\sum_{i=1}^{N}y_i, \ \frac{1}{N}\sigma^2 \Bigg).$$  
Similarly, the conditional distribution of $\sigma^2$, given $(\mu, \tilde{y}_N,\tilde{z}_N, \tilde{\eta}, \tilde{I})$,
is 
$$P(\sigma^2|\mu, \tilde{y}_N,\tilde{z}_N, \tilde{\eta}, \tilde{I}) \propto  \sigma^{-N-2} \exp \left\lbrace -\frac{1}{2\sigma^2}\sum_{i=1}^{N}(y_i-\mu)^2 \right\rbrace.$$
Then, given $(\mu, \tilde{y}_N,\tilde{z}_N, \tilde{\eta}, \tilde{I}), \sigma^2$ has an Inverse-Gamma distribution with shape parameter $N/2$ and scale parameter $\frac{1}{2}\sum_{i=1}^{N}(y_i-\mu)^2$, i.e., 
$$ \sigma^2|\mu, \tilde{y}_N,\tilde{z}_N, \tilde{\eta}, \tilde{I} \sim \mbox{Inverse-Gamma} \Bigg(\frac{N}{2}, \frac{1}{2}\sum_{i=1}^{N}(y_i-\mu)^2\Bigg). $$
The conditional distribution of $\tilde{\eta}$, given $(\tilde{y}_N,\tilde{z}_N, \tilde{\psi}, \tilde{I})$, is
\begin{align*}
P(\tilde{\eta}|\tilde{y}_N,\tilde{z}_N, \tilde{\psi}, \tilde{I}) \propto & \sigma_e^{-N-2} \exp \left\{ -\frac{1}{2\sigma_e^2} \Bigg [  Nz_N^2-2z_N\sum_{i=1}^{N}(\beta_0+\beta_1y_i)
\right. \ \\
&\phantom{=\;\;}\left.\ +\Bigg(\sum_{i=1}^{N-1}z_i+\beta_0+\beta_1y_N \Bigg)^2 + \sum_{i=1}^{N-1} (z_i-\beta_0-\beta_1y_i)^2 \Bigg ] \right\}.
\end{align*}
Then the conditional distribution of $\sigma_e^2$, given $(\beta_0, \beta_1, \tilde{y}_N, \tilde{z}_N,\tilde{\psi}, \tilde{I})$, is 
$$P(\sigma_e^2|\beta_0, \beta_1, \tilde{y}_N, \tilde{z}_N,\tilde{\psi}, \tilde{I}) \propto \sigma_e^{-N-2} \exp \left\lbrace -\frac{1}{2\sigma_e^2}T_4 \right\rbrace,$$
where 
$$T_4 \ = \ Nz_N^2-2z_N\sum_{i=1}^{N}(\beta_0+\beta_1y_i)+\Bigg(\sum_{i=1}^{N-1}z_i+\beta_0+\beta_1y_N \Bigg)^2 + \sum_{i=1}^{N-1} (z_i-\beta_0-\beta_1y_i)^2.$$
So, given $(\beta_0, \beta_1, \tilde{y}_N, \tilde{z}_N,\tilde{\psi}, \tilde{I})$, $\sigma_e^2$ has an Inverse-Gamma distribution with shape parameter $N/2$ and scale parameter $T_4/2$, i.e., 
$$\sigma_e^2|\beta_0, \beta_1, \tilde{y}_N, \tilde{z}_N,\tilde{\psi}, \tilde{I} \sim \mbox{Inverse-Gamma} \Bigg(\frac{N}{2}, \frac{T_4}{2} \Bigg).$$
The conditional distribution of $\beta_0$ given $(\sigma_e^2, \beta_1, \tilde{y}_N, \tilde{z}_N,\tilde{\psi}, \tilde{I})$ is 
$$P(\beta_0|\sigma_e^2, \beta_1, \tilde{y}_N, \tilde{z}_N,\tilde{\psi}, \tilde{I}) \propto \exp\left\lbrace -\frac{1}{2\sigma_e^2}\Bigg[N\beta_0^2+2\Bigg(\beta_1\sum_{i=1}^{N}y_i-t\Bigg)\beta_0 \Bigg] \right\rbrace. $$
This implies that, given $(\sigma_e^2, \beta_1, \tilde{y}_N, \tilde{z}_N,\tilde{\psi}, \tilde{I})$, $\beta_0$ has the following normal distribution 
$$\beta_0|\sigma_e^2, \beta_1, \tilde{y}_N, \tilde{z}_N,\tilde{\psi}, \tilde{I} \sim N \Bigg(-\frac{1}{N}\Bigg(\beta_1\sum_{i=1}^{N}y_i-t\Bigg), \frac{1}{N} \sigma_e^2\Bigg).$$

Similarly, the conditional distribution of $\beta_1$ given $(\sigma_e^2, \beta_0, \tilde{y}_N, \tilde{z}_N,\tilde{\psi}, \tilde{I})$ is 
$$P(\beta_1|\sigma_e^2, \beta_0, \tilde{y}_N, \tilde{z}_N,\tilde{\psi}, \tilde{I}) \propto \exp\left\lbrace -\frac{1}{2\sigma_e^2}\Bigg[\beta_1^2\sum_{i=1}^{N}y_i^2+2T_5\beta_1 \Bigg]\right\rbrace, $$
where 
$$T_5= \sum_{i=1}^{N-1}y_i(\beta_0-z_i)+y_N \Bigg(\beta_0+\sum_{i=1}^{N-1}z_i \Bigg)-\frac{t}{N}\sum_{i=1}^{N}y_i.$$
Then, given $(\sigma_e^2, \beta_0, \tilde{y}_N, \tilde{z}_N,\tilde{\psi}, \tilde{I})$, $\beta_1$ has the following normal distribution $$\beta_1|\sigma_e^2, \beta_0, \tilde{y}_N, \tilde{z}_N,\tilde{\psi}, \tilde{I} \sim N\Bigg(-\frac{T_5}{\sum_{i=1}^{N}y_i^2},
 \frac{1}{\sum_{i=1}^{N}y_i^2}\sigma_e^2\Bigg).$$
After having all the conditional distributions, we use the Gibbs sampler to draw samples from $\pi_a(\tilde{y}_{ns},\tilde{z}_{ns},\tilde{\psi},\tilde{\eta}|\tilde{y}_{s},\tilde{z}_{s},\tilde{I}).$

\section{Evaluating $C(\tilde{y}_N,\tilde{\psi}, \tilde{\eta})$}
The weights in the SIR algorithm are related to the normalization constant  $C(\tilde{y}_N,\tilde{\psi}, \tilde{\eta})$ as shown in Section 2.3. Given  $(\tilde{y}_N,\tilde{\psi}, \tilde{\eta})$, we need to compute  $C(\tilde{y}_N,\tilde{\psi}, \tilde{\eta})$. First, notice that  $C(\tilde{y}_N,\tilde{\psi}, \tilde{\eta})$ can be further rewritten as 
\begin{align}
\nonumber
C(\tilde{y}_N, \tilde{\psi},\tilde{\eta} ) \ = \ & \int_{R_0}^{}N(\sqrt{2\pi}\sigma_e)^{-N} \exp \left\{ -\frac{1}{2\sigma_e^2} \Bigg [ \frac{t^2}{N}-2\frac{t}{N}\sum_{i=1}^{N}\theta_i+\Big(\sum_{i=1}^{N-1}z_i+\theta_N \Big)^2
\right. \ \\
\nonumber
&\phantom{=\;\;}\left.\  + \sum_{i=1}^{N-1} (z_i-\theta_i)^2 \Bigg ] \right\}dz_1dz_2...dz_{N-1} \\ 
\nonumber
= \ & N(\sqrt{2\pi}\sigma_e)^{-N} \exp \left\lbrace -\frac{1}{2\sigma_e^2} \Bigg [ \frac{t^2}{N}-2\frac{t}{N}\sum_{i=1}^{N}\theta_i \Bigg ] \right\rbrace \\
\nonumber
& \ \times  \int_{R_0}^{} \exp \left\lbrace -\frac{1}{2\sigma_e^2} \Bigg[ \Big( \sum_{i=1}^{N-1}z_i+\theta_N \Big)^2+\sum_{i=1}^{N-1} (z_i-\theta_i)^2 \Bigg] \right\rbrace dz_1dz_2...dz_{N-1}.
\end{align}

Let $g(\tilde{z}_N,\tilde{y}_N,\tilde{\psi}, \tilde{\eta})$ denote 
$$\exp \left\lbrace -\frac{1}{2\sigma_e^2} \Bigg[ \Big( \sum_{i=1}^{N-1}z_i+\theta_N \Big)^2+\sum_{i=1}^{N-1} (z_i-\theta_i)^2 \Bigg] \right\rbrace. $$
We also define $G(\tilde{y}_N,\tilde{\psi}, \tilde{\eta})$ as
\begin{align*}
\nonumber
G(\tilde{y}_N,\tilde{\psi}, \tilde{\eta}) \ = \ &  \int_{R_0}^{} \exp \left\lbrace -\frac{1}{2\sigma_e^2} \Bigg[ \Big( \sum_{i=1}^{N-1}z_i+\theta_N \Big)^2+\sum_{i=1}^{N-1} (z_i-\theta_i)^2 \Bigg] \right\rbrace dz_1dz_2...dz_{N-1} \\
 = \ &  \int_{R_0}^{}g(\tilde{z}_N,\tilde{y}_N,\tilde{\psi}, \tilde{\eta})dz_1dz_2...dz_{N-1}.
\end{align*}
We separate region $R_0$ into two parts, $R_1$ and $R_2$:
$$R_1 \ = \ \Bigg\{ (z_1, z_2,...,z_{N-1}| -\frac{t}{N} \leq z_i \leq \frac{t}{n}-\frac{t}{N})  \Bigg\}, $$
$$R_2 \ = \ \Bigg\{ (z_1, z_2,...,z_{N-1}| \frac{t}{N}-\frac{t}{n} \leq \sum_{i=1}^{N-1}z_i \leq \frac{t}{N}) \Bigg\}.$$
Then we have 
\begin{align*}
\nonumber
G(\tilde{y}_N,\tilde{\psi}, \tilde{\eta}) \ = \ &  \int_{R_0}^{}g(\tilde{z}_N,\tilde{y}_N,\tilde{\psi}, \tilde{\eta})dz_1dz_2...dz_{N-1} \\
\ = \ & \int_{R_1}^{}I_{R_2}g(\tilde{z}_N,\tilde{y}_N,\tilde{\psi}, \tilde{\eta})dz_1dz_2...dz_{N-1} \\
\ = \ & \int_{R_1}^{}(I_{R_2}C_0)\frac{1}{C_0}g(\tilde{z}_N,\tilde{y}_N,\tilde{\psi}, \tilde{\eta})dz_1dz_2...dz_{N-1},
\end{align*}
where $C_0$ is defined as 
\begin{align*}
\nonumber
C_0 \ = \ &  \int_{R_1}^{}g(\tilde{z}_N,\tilde{y}_N,\tilde{\psi}, \tilde{\eta})dz_1dz_2...dz_{N-1} \\
= \ &\int_{R_1}^{} \exp \left\lbrace -\frac{1}{2\sigma_e^2} \Bigg[ \Big( \sum_{i=1}^{N-1}z_i+\theta_N \Big)^2+\sum_{i=1}^{N-1} (z_i-\theta_i)^2 \Bigg] \right\rbrace dz_1dz_2...dz_{N-1}.
\end{align*}
From the above definition, we see that $\frac{1}{C_0}g(\tilde{z}_N,\tilde{y}_N,\tilde{\psi}, \tilde{\eta})$ is a multivariate normal density restricted to $R_1$ and $C_0$ is the corresponding normalization constant. Now consider the following integral,
$$\int_{R_1}^{}I_{R_2}\frac{1}{C_0}g(\tilde{z}_N,\tilde{y}_N,\tilde{\psi}, \tilde{\eta})dz_1dz_2...dz_{N-1}.$$

One can compute this integral by drawing samples from the multivariate normal density $\frac{1}{C_0}g(\tilde{z}_N,\tilde{y}_N,\tilde{\psi}, \tilde{\eta})$ restricted to $R_1$ and counting how many samples fall in $R_2$. It is our experience that this proportion usually is close to 1. Notice that $g(\tilde{z}_N,\tilde{y}_N,\tilde{\psi}, \tilde{\eta})$ is the kernel of the multivariate normal distribution with mean $\tilde{\mu}'$ and variance $\tilde{\Sigma}'$ where

$$
\tilde{\mu}'=\begin{pmatrix}
\mu_{1}' \\
\mu_{2}'  \\
...  \\
\mu_{N-1}' \\
\end{pmatrix} =\begin{pmatrix}
\frac{1}{N}[(N-1)\theta_{1}-\theta_{2}-...-\theta_{N-1}-\theta_N] \\
\frac{1}{N}[(N-1)\theta_{2}-\theta_{1}-...-\theta_{N-1}-\theta_N]  \\
...  \\
\frac{1}{N}[(N-1)\theta_{N-1}-\theta_{1}-...-\theta_{N-2}-\theta_N] \\
\end{pmatrix} _{(N-1) \times 1}
$$
and
\begin{equation}
\tilde{\Sigma'}^{-1}=\frac{1}{\sigma_e^2}\begin{pmatrix}
2 & 1 & ... & 1 \\
1 & 2 & ... & 1 \\
. & . & ... & . \\
1 & 1 & ... & 2 \\
\end{pmatrix}_{(N-1) \times (N-1)}.
\end{equation}

We still need to evaluate $C_0$, the normalization constant for $g(\tilde{z}_N,\tilde{y}_N,\tilde{\psi}, \tilde{\eta})$, which can be rewritten as 
\begin{align*}
\nonumber
C_0 \ = \  &\int_{R_1}^{} \exp \left\lbrace -\frac{1}{2\sigma_e^2} \Bigg[ \Big( \sum_{i=1}^{N-1}z_i+\theta_N \Big)^2+\sum_{i=1}^{N-1} (z_i-\theta_i)^2 \Bigg] \right\rbrace dz_1dz_2...dz_{N-1}\\
 = \  & \int_{R_1}^{} \Bigg[ \exp \left\lbrace -\frac{1}{2\sigma_e^2} \Bigg[ \theta_N^2+ \sum_{i=1}^{N-1}\theta_i^2-\Bigg(\sum_{i=1}^{N-1}\mu'_i \Bigg)^2-\sum_{i=1}^{N-1}(\mu'_i)^2 \Bigg] \right\rbrace (2\pi)^{\frac{N-1}{2}}\sqrt{|\tilde{\Sigma}'|} \Bigg] \\
& \ \times \frac{1}{(2\pi)^{\frac{N-1}{2}}\sqrt{|\tilde{\Sigma}'|}} \exp \left\lbrace -\frac{1}{2} \Big[{(\tilde{z}_{N-1}-\tilde{\mu'})}' {\Sigma'}^{-1} (\tilde{z}_{N-1}-\tilde{\mu'}) \Big] \right\rbrace dz_1dz_2...dz_{N-1} \\
= \ & \Bigg[ \exp \left\lbrace -\frac{1}{2\sigma_e^2} \Bigg[ \theta_N^2+ \sum_{i=1}^{N-1}\theta_i^2-\Bigg(\sum_{i=1}^{N-1}\mu'_i \Bigg)^2-\sum_{i=1}^{N-1}(\mu'_i)^2 \Bigg] \right\rbrace (2\pi)^{\frac{N-1}{2}}\sqrt{|\tilde{\Sigma}'|} \Bigg] \\
& \ \times \int_{R_1}^{} \frac{1}{(2\pi)^{\frac{N-1}{2}}\sqrt{|\tilde{\Sigma}'|}} \exp \left\lbrace -\frac{1}{2} \Big[{(\tilde{z}_{N-1}-\tilde{\mu'})}' {\Sigma'}^{-1} (\tilde{z}_{N-1}-\tilde{\mu'}) \Big] \right\rbrace dz_1dz_2...dz_{N-1}.  
\end{align*}
Notice that 
$$\int_{R_1}^{} \frac{1}{(2\pi)^{\frac{N-1}{2}}\sqrt{|\tilde{\Sigma}'|}} \exp \left\lbrace -\frac{1}{2} \Big[{(\tilde{z}_{N-1}-\tilde{\mu'})}' {\Sigma'}^{-1} (\tilde{z}_{N-1}-\tilde{\mu'}) \Big] \right\rbrace dz_1dz_2...dz_{N-1}$$
is the probability of a multivariate normal distribution over region $R_1$. This is a standard problem and one can use the algorithm proposed by Genz (1992) to compute this probability. By putting everything together, we see how $C(\tilde{y}_N,\tilde{\psi}, \tilde{\eta})$ is evaluated, i.e.,
\begin{align*}
\nonumber
C(\tilde{y}_N,\tilde{\psi}, \tilde{\eta}) \ = \  &  N(\sqrt{2\pi}\sigma_e)^{-N} \exp \left\lbrace -\frac{1}{2\sigma_e^2} \Bigg [ \frac{t^2}{N}-2\frac{t}{N}\sum_{i=1}^{N}\theta_i \Bigg ] \right\rbrace \\
\nonumber
& \ \times  \int_{R_0}^{} \exp \left\lbrace -\frac{1}{2\sigma_e^2} \Bigg[ \Big( \sum_{i=1}^{N-1}z_i+\theta_N \Big)^2+\sum_{i=1}^{N-1} (z_i-\theta_i)^2 \Bigg] \right\rbrace dz_1dz_2...dz_{N-1} \\
= \ & N(\sqrt{2\pi}\sigma_e)^{-N} \exp \left\lbrace -\frac{1}{2\sigma_e^2} \Bigg [ \frac{t^2}{N}-2\frac{t}{N}\sum_{i=1}^{N}\theta_i \Bigg ] \right\rbrace \\
 \nonumber
 & \ \times C_0 \int_{R_1}^{}I_{R_2}\frac{1}{C_0}g(\tilde{z}_N,\tilde{y}_N,\tilde{\psi}, \tilde{\eta})dz_1dz_2...dz_{N-1} \\
 = \ & N(\sqrt{2\pi}\sigma_e)^{-N} \exp \left\lbrace -\frac{1}{2\sigma_e^2} \Bigg [ \frac{t^2}{N}-2\frac{t}{N}\sum_{i=1}^{N}\theta_i \Bigg ] \right\rbrace \\
 \nonumber
 & \ \times \Bigg[ \exp \left\lbrace -\frac{1}{2\sigma_e^2} \Bigg[ \theta_N^2+ \sum_{i=1}^{N-1}\theta_i^2-\Bigg(\sum_{i=1}^{N-1}\mu'_i \Bigg)^2-\sum_{i=1}^{N-1}(\mu'_i)^2 \Bigg] \right\rbrace (2\pi)^{\frac{N-1}{2}}\sqrt{|\tilde{\Sigma}'|} \Bigg] 
  \end{align*}
 \begin{align*}
 & \qquad \qquad \qquad \times \int_{R_1}^{} \frac{1}{(2\pi)^{\frac{N-1}{2}}\sqrt{|\tilde{\Sigma}'|}} \exp \left\lbrace -\frac{1}{2} \Big[{(\tilde{z}_{N-1}-\tilde{\mu'})}' {\Sigma'}^{-1} (\tilde{z}_{N-1}-\tilde{\mu'}) \Big] \right\rbrace dz_1dz_2...dz_{N-1}\\
 & \qquad \qquad \qquad \times  \int_{R_1}^{}I_{R_2}\frac{1}{C_0}g(\tilde{z}_N,\tilde{y}_N,\tilde{\psi}, \tilde{\eta})dz_1dz_2...dz_{N-1}.
\end{align*}

In the SIR algorithm the computational burden is to compute this normalization constant, $C(\tilde{y}_N,\tilde{\psi}, \tilde{\eta})$. This has to be done at each Gibbs sampler step. Specifically, at each Gibbs sampler step, we need to draw samples from $g(\tilde{z}_N,\tilde{y}_N,\tilde{\psi}, \tilde{\eta})$ using another Gibbs sampler to compute
$$\int_{R_1}^{}I_{R_2}\frac{1}{C_0}g(\tilde{z}_N,\tilde{y}_N,\tilde{\psi}, \tilde{\eta})dz_1dz_2...dz_{N-1}.$$
We also need to calculate the $(N-1)$ dimensional multivariate normal probability 
$$\int_{R_1}^{} \frac{1}{(2\pi)^{\frac{N-1}{2}}\sqrt{|\tilde{\Sigma}'|}} \exp \left\lbrace -\frac{1}{2} \Big[{(\tilde{z}_{N-1}-\tilde{\mu'})}' {\Sigma'}^{-1} (\tilde{z}_{N-1}-\tilde{\mu'}) \Big] \right\rbrace dz_1dz_2...dz_{N-1}$$
at each Gibbs sampler step. It is our experience that $C(\tilde{y}_N,\tilde{\psi}, \tilde{\eta})$ usually is a very small number. For this purpose, we have developed a parallel computing algorithm for SIR in a high-performance cluster environment.
\end{appendices}

%%%%%%%%%%%%%%%%%%%%%%%%%%%%%%%%%%%%%%%%%%%%%%%%%%%%%%%%%%%%%%%%%%%%%%%%%%%%%%%%%%%%%%%%%%%%%%%%%%%%%%%%%%%%%%%%%%%%%%%%%%%%%%

\noindent
{\bf Acknowledgments}\\
This work is a presentation to honor JNK Rao on his  eightieth birthday.
This research was supported by a grant from the Simons Foundation (\#353953, Balgobin Nandram).

%%%%%%%%%%%%%%%%%%%%%%%%%%%%%%%%%%%%%%%%%%%%%%%%%%%%%%%%%%%%%%%%%%%%%%%%%%%%%%%%%%%%%%%%%%%%%%%%%%%%%%%%%%%%%%%%%%%%%%%%%%%%%%%		 

%\newpage

\begin{mybibliography}{}
		
\bibitem[]{} Chambers, R., Dorfman, A., and Wang, S. (1998). 
\newblock  { Limited Information Likelihood Analysis of
	Survey Data}.
\newblock  {\it Journal of the Royal Statistical Society}, Series B, \textbf{60}, 397-411.

\bibitem[]{} Gelfand, A. E. and Smith, A. F. M. (1990). 
\newblock {Sampling-based Approaches to Calculating Marginal
	Densities}.
\newblock  {\it Journal of the American Statistical Association}, \textbf{85}, 398-409.

\bibitem[]{} Genz, A. (1992).
\newblock {Numerical Computation of Multivariate Normal Probabilities}.
\newblock {\it Journal of Computational
	and Graphical Statistics}, \textbf{1}, 141-149.

\bibitem[]{} Krieger, A. M. and Pfeffermann, D. (1992). 
\newblock { Maximum Likelihood Estimation From Complex Sample
	Surveys}.
\newblock  {\it Survey Methodology}, \textbf{18(2)}, 225-239.
		
\bibitem[]{} Ma, J. (2010). 
\newblock    { Contributions to Numerical Formal Concept Analysis, Bayesian Predictive Inference
	and Sample Size Determination}.
\newblock    {\it Unpublished Ph.D. Thesis, Case Western Reserve University}.
			
\bibitem[]{} Malec, D., Davis,W., and Cao, X. (1999).
\newblock    { Model-based Small Area Estimates of Overweight Prevalence
	Using Sample Selection Adjustment}. 
\newblock    {\it Statistics in Medicine},  \textbf{18}, 3189-3200. 

\bibitem[]{} Nandram, B., Bhatta, D., Bhadra, D., and Shen, G. (2013).
\newblock   { Bayesian Predictive Inference of a Finite Population Proportion under Selection Bias}.
\newblock   {\it Statistical Methodology}, \textbf{11}, 1-21.

\bibitem[]{} Nandram, B. and Choi, J. W. (2010).
\newblock   {A Bayesian Analysis of Body Mass Index Data from 
            Small Domains Under Nonignorable Nonresponse and Selection}.
\newblock   {\it Journal of the American Statistical Association}, \textbf{105}, 120-135.

\bibitem[]{} Pfeffermann, D. and Sverchkov, M. (2009).
\newblock    {Inference Under Informative Sampling}.
\newblock    {\it Handbook
	on Statistics, Sample Surveys: Inference and Analysis}, \textbf{29(B)}, 455-487.

\bibitem[]{} Pfeffermann, D., Da Silva Moura, F., and Do Nascimento Silva, P. (2006). 
\newblock   {Multi-Level Modelling
	Under Informative Sampling}.
\newblock   {\it Biometrika}, \textbf{93}, 943-959. 

\bibitem[]{} Si, Y., Pillai, N. S., and Gelman, A. (2015).
\newblock    {Bayesian Nonparametric Weighted Sampling Inference}.
\newblock {\it Bayesian Analysis}, \textbf{10(3)}, 605-625.

\bibitem[]{} Smith, A. F. M. and Gelfand, A. E. (1992).
\newblock    {Bayesian Statistics without Tears: A Sampling Resampling
	Perspective}.
\newblock {\it The American Statistician}, \textbf{85}, 84-88.
		
\bibitem[] {} Zangeneh, S. and Little, R. (2015). 
\newblock {Bayesian Inference for the Finite Population Total from a Heteroscedastic
	Probability Proportional to Size Sample}.
\newblock  {\it Journal of Survey Statistics and Methodology}, \textbf{3}, 162-192.

\end{mybibliography}

\end{document}